# Enhancement of the superconducting critical temperature realized in the La–Ce–H system at moderate pressures

Wuhao Chen,[1] Xiaoli Huang,[1,*] Dmitrii V. Semenok,[2] Su Chen,[1] Kexin Zhang,[1] Artem R. Oganov[2] and Tian Cui[3,1,*]

[1] State Key Laboratory of Superhard Materials, College of Physics, Jilin University, Changchun 130012, China
[2] Skolkovo Institute of Science and Technology, Skolkovo Innovation Center, Bolshoy Boulevard 30, bldg. 1, Moscow, 121205, Russia
[3] School of Physical Science and Technology, Ningbo University, Ningbo 315211, China

*Corresponding authors' e-mails: huangxiaoli@jlu.edu.cn, cuitian@nbu.edu.cn

**Abstract**

Covalent and ionic polyhydrides have become the two main camps in searching for the high-temperature superconductors under pressure. They have been considered as important platforms for exploring ternary or multiple hydrides in order to further increase the $T_c$ or decrease the stabilization pressure. In this work, we successfully synthesized ternary hexagonal La–Ce polyhydrides stable in the pressure range of 95–130 GPa by laser-heating the La–Ce alloy (initial ratio La:Ce = 2.5–3.5:1) in ammonia borane. Superconductivity at 176 K was strikingly preserved to about 100 GPa. The extrapolated upper critical field $H_{c2}(0)$ reached 216 T at 100 GPa, the highest value among the synthesized polyhydrides. We also performed the contrast experiments and stabilized binary high-temperature superconducting $LaH_x$ with $T_c$ ~103 K at 78 GPa. In the pressure range of 95–130 GPa, the ternary hexagonal La–Ce–H system exhibits higher $T_c$ than the binary La–H system, with the maximum difference of 100 K, and the compounds of both systems were synthesized at the same pressure and temperature conditions. These results clearly indicate that the discovered La–Ce–H system not only enriches the high-temperature superconducting hydrides but also realizes high-$T_c$ at moderate pressures.

**Introduction**

Recent discoveries of hydrogen-based superconductors, such as $H_3S$ [1-5], $LaH_{10}$ [6-9], $CeH_9$ and $CeH_{10}$ [10-12], $YH_6$ and $YH_9$ [13-15], have inspired enormous interest in multiple areas. Binary high-temperature superconducting (HTSC) hydrides are basically divided into two categories: covalent and ionic. The first theoretically [2] and experimentally [1] discovered hydride $H_3S$ belongs to the class of covalent polyhydrides, which requires covalent bonding between hydrogen and light nonmetal elements located in the upper right corner of the periodic table. The second type is mainly alkaline and rare earth metal polyhydrides with sodalite-like clathrate structures [16,17], currently leading the relevant research. Besides, the layered hydrides with graphene-like hydrogen net are predicted to have a great potential as high-temperature superconductors [18]. Although the superconducting records have been constantly refreshing, obtaining high-temperature superconducting phases at moderate pressures is still misty. The attention has been turned to ternary hydrides. Sun et al. theoretically predicted the ionic metal hydride $Li_2MgH_{16}$ with the critical temperature $T_c$ of about 473 K at 250 GPa [19], indicating a huge possibility for HTSC in ternary systems. The recent theoretical work proposed a series of $ABH_8$-type hydrides with HTSC properties [20-23], especially $LaBeH_8$, dynamically stable down to 20 GPa with high $T_c \sim 185$ K [23]. In contrast to rich theoretical studies, ternary hydrides have more research space in experiment, however, they remained elusive because of uncontrollable and complex synthesis.

Lanthanum [6-9,24-27] and cerium superhydrides [10-12,28-30] have been well studied before, and sharing the same HTSC structures ($Fm\bar{3}m$ and $P6_3/mmc$). Lanthanum superhydrides show higher $T_c$, whereas the synthesis conditions of cerium superhydrides are more favorable. As neighbors in the periodic table, La and Ce have very close atom radii and electronegativity and can form continuous and homogeneous solid solution [31,32]. Thus, it is possible that using a La–Ce alloy as the initial reactant, we may get ternary hydrides with structures similar to those of the binary La–H and Ce–H. A recent experimental work revealed that, even belonging to the different periods, La and Y can be partly replaced by each other and form HTSC hydrides with almost the same structure as their binary hydrides [33]. Alloys may exhibit superior properties compared with their individual components, and these properties can be adjusted by the relative content. Alloys of Ce also show irregularities under pressure due to the *f* electron delocalization and have the tendency to form mixed-valence states already at ambient pressure [34,35]. In this work, we have synthesized ternary HTSC *hcp*-(La,Ce)H$_9$-

$_{10}$, which are stable in the pressure range of 95–130 GPa and has the same structure as binary polyhydrides. In the explored pressure range, the La–Ce–H system displayed the highest $T_c$ compared with the binary La–H and Ce–H systems. Superconductivity in the La–Ce–H system at about 176 K was strikingly preserved to ~100 GPa with a quite high figure of merit $S$=1.62 among the synthesized polyhydrides.

**Results and Discussion**

As the initial reactant, La–Ce alloys were prepared using the multitarget magnetron sputtering with the atomic ratio of La: Ce ratio of 2.5–3.5 for typical experimental runs (Table S1 and Fig. S1). Before loading into the diamond anvil cells (DACs), the alloys were characterized using the scanning electron microscopy with energy dispersive X-ray spectroscopy (SEM+EDX) (Figs. S11–S13, and S16), X-ray diffraction (XRD) (Fig. S2) and electrical measurements (Figs. S14 and S15), which indicated good homogeneity. The La: Ce ratios of the alloys in runs #1,2,3,7 and 8 should be close to 3, because these alloys were prepared in similar conditions (Fig. S1). After synthesizing La–Ce–H compounds by laser-heating at specific pressures, we conducted the electrical measurements and plotted the typical results (Fig. 1). To synthesize high-$T_c$ phases at low pressures, we heated DACs #2 and #9 at 113 GPa and 120 GPa, obtaining $T_c$s of 175 K and 190 K, respectively. Moreover, the $T_c$ can be preserved at a value of 155 K at 95 GPa (DAC #2) and 180 K at 104 GPa (DAC #9). To make a comparison with the reported LaH$_{10}$, we heated DAC #6 at 152 GPa, and observed the resistance drop starting from 187 K. $T_c$ slightly decreased with further compression to 156 GPa and subsequently 162 GPa, which is different from the behavior of $C2/m$-LaH$_{10}$ [24]. Noteworthy, the pressure scale (diamond Raman edge) used in this study gives a higher value than that of the hydrogen vibron by ~18 GPa [24]. DACs #3 and #5 were both laser-heated at about 130 GPa. However, $T_c$ in DAC #5 reached 188 K at 132 GPa, higher than $T_c$ achieved in DAC #3 (163 K at 131 GPa), possibly because of the small difference in the La content in the La–Ce alloy used. Upon decreasing the pressure in DAC #5 by 5 GPa, two apparent steps appeared in the temperature dependence of the resistance. Higher $T_c$, decreasing gradually along with the pressure, dropped to 132 K at 104 GPa. In contrast, lower $T_c$ phase (~37 K, at the resistance close to zero) was robust during decompression from 127 GPa (Fig. 1b) to at least 80 GPa (Fig. 1c). Further decompression of the sample in DAC #9 from 123 to 101 GPa showed the tendency similar to that of the high-$T_c$ phase in DACs #2 and #5 (Fig. 1). The resistance of the La–

Ce–H samples increased significantly when the pressure decreased, and the width of the transition increased about 2.5 times from 104 GPa to 101 GPa (Fig. 1c), which indicates the instability of the high-$T_c$ phase in this pressure range.

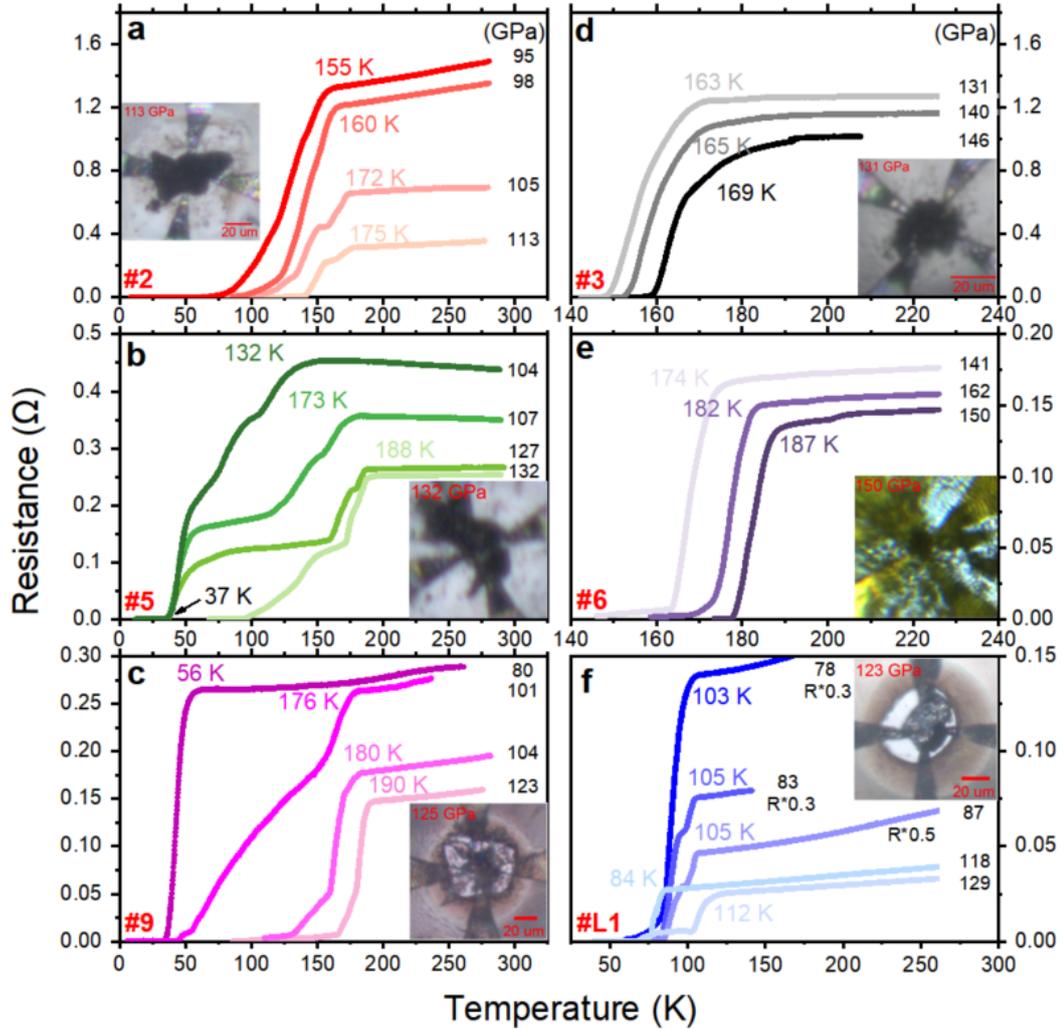

**FIG. 1.** Characterization of the superconducting transition using electrical resistance measurements at selected pressures for the typical runs. DAC #L1 is the La-H sample, and the others are La–Ce–H samples. Insets are the photographs of the sample chambers. The residual resistances are shown in Fig. S3.

To further confirm high-temperature superconductivity, an external magnetic field was applied to DAC #2 for studying the upper critical field $H_{c2}$ at 100 GPa (Fig. 2a). Fitting using the Werthamer-Helfand-Hohenberg (WHH) [36] model gives $H_{c2}(0)$ = 216 T, the highest value reported for superconducting hydrides so far. The inhomogeneity of the synthesized sample may introduce more defects, which enhances the flux pining. In this case, the external magnetic field would distribute unevenly in the sample and finally contribute to the large value of $H_{c2}(0)$. An important question in studies of hydride superconductivity is the presence or absence of anisotropy in the behavior of samples in an external magnetic field. To investigate possible anisotropy, we applied a magnetic field parallel and perpendicular to the culet in DAC #9, keeping the sample exactly in the center of the field (Fig. 2b). The upper critical field was measured first in the perpendicular position at 125 GPa, showing $H_{c2}(0)$ = 146 T. After DAC #9 was taken out of the cryostat and rotated by 90°, the pressure increased by 2 GPa, thus both the onset resistance and $T_c$ decreased slightly. However, $H_{c2}(0)$ increased to 163 T. Therefore, this La–Ce–H sample may have a texture, which leads to the difference in the upper critical magnetic fields: $H_{c2\parallel} > H_{c2\perp}$. This is the first such experimental study of anisotropy for superhydrides in DACs.

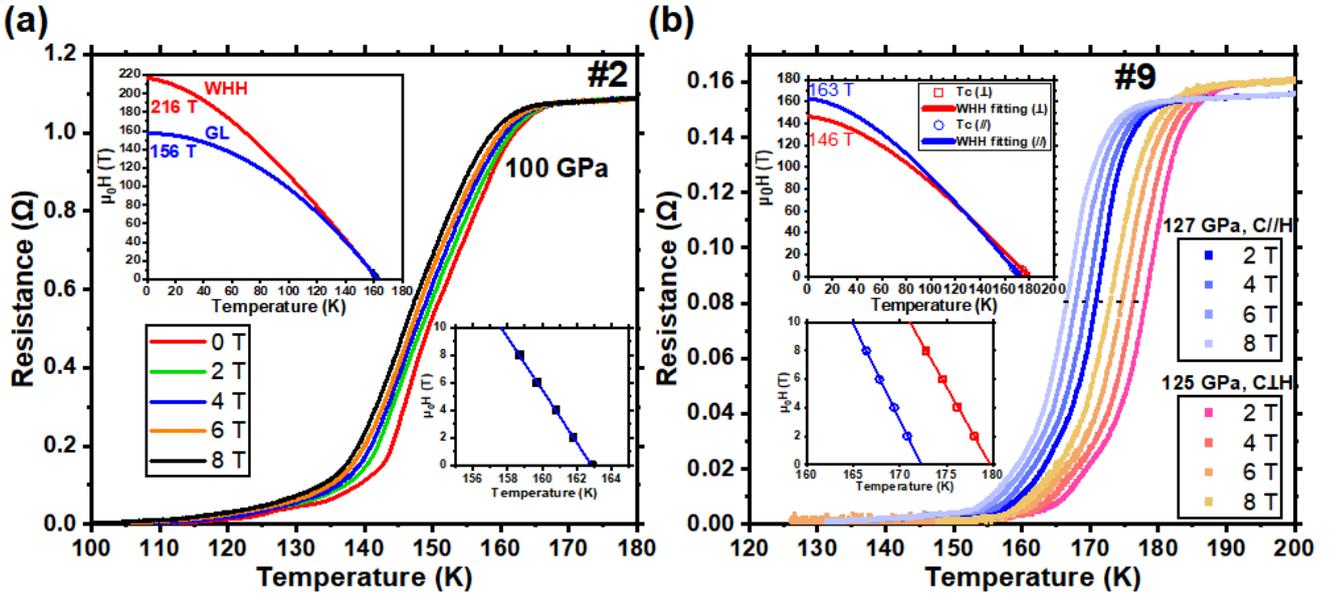

**FIG. 2.** Temperature dependences of the electrical resistance near the superconducting transition in external magnetic fields. (a) DAC #2. Insets show the fitting with WHH and GL models and the enlarged part of the fitting. (b) DAC #9. C∥H and C⊥H indicate the cases of the field parallel and perpendicular to the culet, respectively. Insets show the fitting with WHH model and the enlarged part of the fitting.

Compared with the phases studied in our previous work on the binary Ce–H system [10], here we successfully synthesized the ternary La–Ce–H compounds with higher $T_c$ and $H_{c2}$. However, we found no work reporting the electrical measurements in the binary La–H system at pressures lower than 120 GPa. To fill this gap, we performed the electrical experiments in the similar pressure range with the ternary system. For the La–H sample in DAC #L1, an obvious resistance drop appeared at 84 K after the first laser-heating at 123 GPa, with rather low $H_{c2}(0)$ ~ 24.5 T (Fig. S17 and S20). Further increasing the pressure and laser heating the sample in DAC #L1 led to another superconducting transition with higher $T_c$ of 112 K at 129 GPa (Fig. S21). During the decompression of the DAC #L1, circular cracks appeared on the bevel of the diamond (Fig. S18 and S19). Luckily, the electrodes kept interconnected upon decompression to 78 GPa, and the sample still had the $T_c$ of about 103 K (Fig. 1f). Considering $T_c$ of 190 K at 123 GPa in DAC #9 for the La–Ce–H sample, the maximum difference of $T_c$ between the La-Ce-H and La-H samples is about 100 K.

Currently, there is no effective way to reduce the plastic deformation of the insulating gasket. Considering that the circular cracks are related to the cleavage feature and the anisotropy of hardness peculiar to single-crystal diamond (SCD) [37], we tested the nano-polycrystalline diamond (NPD) then [38]. NPD consists of randomly oriented fine diamond nanocrystals and has higher Knoop hardness than SCD [39]. Besides, NPD has been well studied for the ultrahigh-pressure measurements [40-43]. Here, we used NPD for the decompression process. Additionally, it is expectable that the longitudinal cleavage plane (Fig. S4b) arising inside the SCD mainly because of the hydrogen diffusion can be effectively suppressed. We combined a flat NPD with a 200 μm culet and a beveled SCD (200 μm to 100 μm in 8.5°) in DAC #L2 (Fig. S22). The pressure dropped from 113 GPa to 111 GPa after the laser-heating, and a very clear transition was detected at 80 K (Fig. S23). Upon further decompression, $T_c$ can be tracked to 92 K at 77 GPa (Fig. S24).

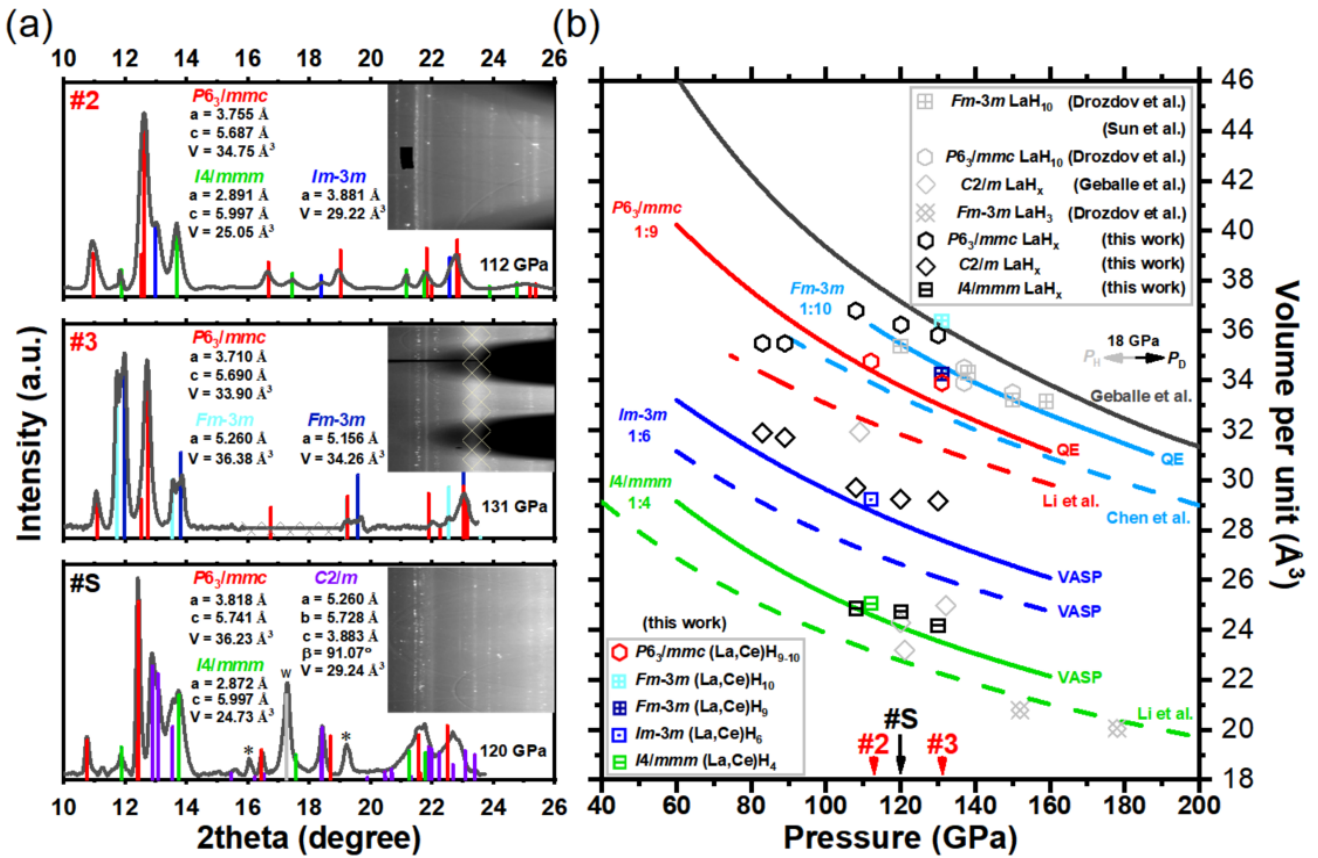

**FIG. 3.** Synchrotron X-ray diffraction (0.6199 Å) analysis of the synthesized hydrides. (a) Peaks indexing

for the La–Ce–H samples in the DACs #2, 3 and the La–H sample in the DAC #S. Insets show the integrated diffraction patterns. Wide diffraction band of an impurity that located on the seat surface in DAC #3 is masked by gridlines (Fig. S9). (b) Pressure dependence of the unit cell volume of different polyhydrides. The experimental results for the La–Ce–H and La–H systems are shown in color and black, respectively. Gray symbols show literature data for the synthesized La–H phases [7,9,24]. Solid and dashed lines indicate the *P–V* relation of La–H and Ce–H phases, respectively. "VASP" marks the equation of state (EoS) calculated using the VASP code (PBE GGA), and "QE" marks the EoS calculated using the Quantum ESPPESSO code (PAW PBE).

To reveal the real phases and structures determining superconductivity, we performed the synchrotron X-ray diffraction (XRD) measurements on the electrically characterized La–Ce–H samples in DACs #2 and #3, (Fig. 3a, S5, S6 and S8) and newly prepared La–H samples in DAC #S (Fig. 3a and Fig. S26). The data was collected from three individual La–Ce–H electrical DACs at different pressures, and from La–H DAC #S during the decompression. All the superconducting phases in binary La–H and Ce–H system should be considered first of all, and the analysis of the XRD patterns indicated the existence of the hexagonal $P6_3/mmc$ phase. The hydrogen content in the obtained hexagonal phase can be estimated from the unit cell volume. We plotted our *P–V* data together with the calculated or reported equations of state (EoS) of binary Ce–H and La–H hydrides for comparison (Fig. 3b). First, all the EoS curves of CeH$_x$ (dashed lines) are located below those of LaH$_x$ (solid lines) for the same structures with equal hydrogen content. Synthesized *hcp*-(La,Ce)H$_x$ (red hexagons) shows smaller unit cell volume than *hcp*-LaH$_x$ (black hexagons), and they are both located between the experimentally results for $P6_3/mmc$-CeH$_9$ [11] and $Fm\bar{3}m$ -LaH$_{10}$ [9]. The exact hydrogen content cannot be determined, mainly because of an uncertainty in the pseudopotentials of Ce and La [33], and the difference between the pressure scales [7,24,44]. We tentatively supposed the HTSC phase for the La–Ce–H system to be $P6_3/mmc$-(La,Ce)H$_{9-10}$. Considering that the electrical resistance and XRD measurements are not performed on the same DAC for the binary La–H system, $T_c$s observed in DAC #L1 cannot be directly distinguished from that of the mixed phases $P6_3/mmc$-LaH$_x$, and $C2/m$-LaH$_x$ and $I4/mmm$-LaH$_x$ in DAC #S. However, this does not affect the conclusion that the ternary hexagonal La–Ce–H system exhibits higher $T_c$ than the binary La–H system, both of which were synthesized at the same high-pressure and high-temperature conditions.

Besides the hexagonal structure, two cubic polyhydrides were discovered in the La–Ce–H system at 131 GPa in DAC #3 (Fig. 3a), which can be easily distinguished from binary $Fm\bar{3}m$-LaH$_{10}$ because of much lower $T_c$ (Fig. 1d). Previously, Geballe et al. have studied the mixture of La and pure hydrogen by laser heating it to 2000 K at similar pressures of 120–130 GPa [9]. However, they came to different results: the hydrogen content of the synthesized $C2/m$-LaH$_x$ was much lower (gray rhombuses, Fig. 3b) than in our experiment. $I4/mmm$-(La,Ce)H$_x$, $Im\bar{3}m$-(La,Ce)H$_x$ and $C2/m$-LaH$_x$ were also experimentally discovered in this work. The present XRD study shows that the La and Ce atoms possibly form a solid solution in the structure of the ternary hexagonal La–Ce–H hydride, as it was observed for the La-Y-H [33] and La-Nd-H systems [45].

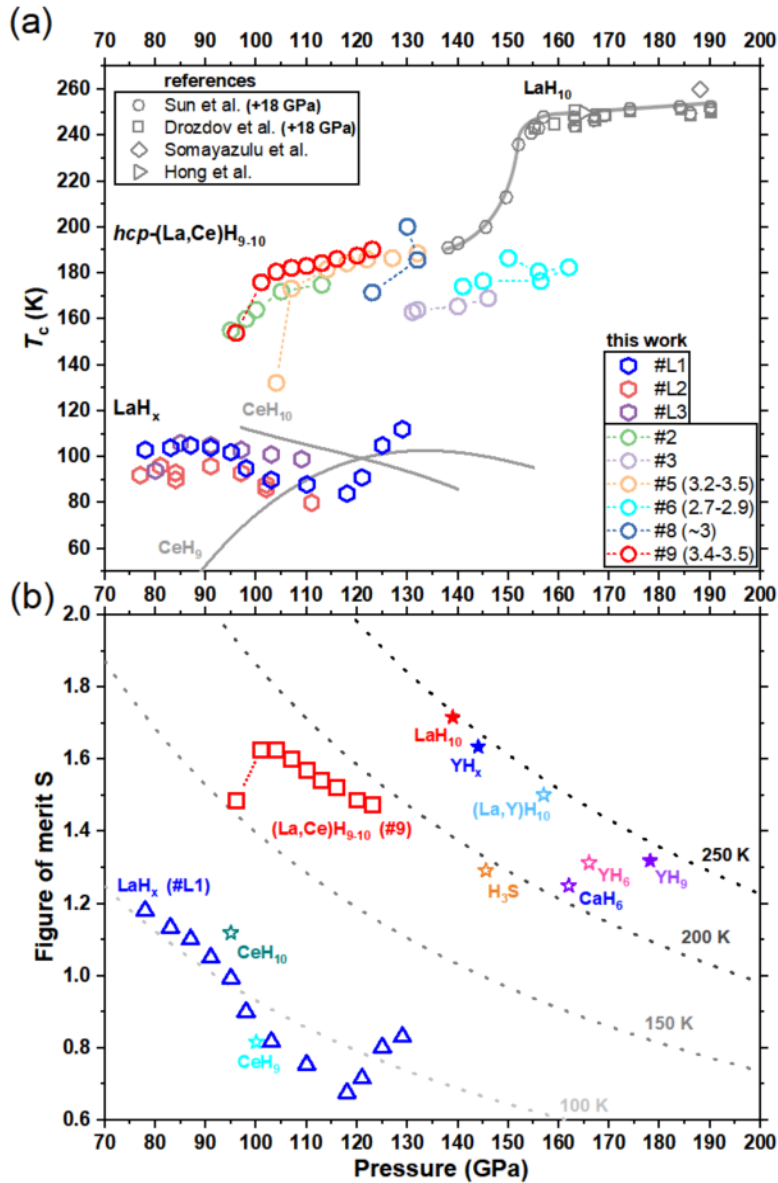

**FIG. 4.** Superconducting critical temperature $T_c$ and the figure of merit S at different pressures. (a) $T_c$–$P$ relationship of the binary La–H and Ce–H and ternary La–Ce–H systems. Gray solid lines indicate the trends. The data from Sun et al. (gray circles) and Drozdov et al. (gray squares) was obtained using the Eremets' pressure scale based on the Raman spectra of hydrogen vibration, which is different from Akahama's pressure scale used in our experiments [46]. Therefore, we added a pressure offset of 18 GPa to make a better comparison with our results. (b) $S$-$P$ diagram of $hcp$-(La,Ce)H$_{9-10}$ and LaH$_x$, including the experimentally highest $S$ value of the well-known superconducting polyhydrides [5,10,14,15,17,24,33,44]. Solid symbols use the pressure scale of the hydrogen vibration. Dashed lines represent isotherms.

The $T_c$–$P$ trends of the binary and ternary systems are plotted in Fig. 4a. $T_c$ of $hcp$-(La,Ce)H$_{9-10}$ decreases with a slope of about 0.6 K/GPa from 130 GPa to 105 GPa, with the maximum value of 200 K at 130 GPa in DAC #8 (Fig. S15). We deem it necessary to clarify several points to exclude the possibility that the transition comes from the binary La–H or Ce–H systems. First, all the SEM+EDX, XRD and $R$–$T$ measurements revealed good quality of the La–Ce alloy at ambient conditions. After the compression of the samples to a certain pressure, electrical measurements showed a sharp transition at a temperature obviously lower than $T_c$ of pure La (Fig. S14 and S15) [47]. This indicates that the introduction of Ce atoms actually affects $T_c$. Second, the superconducting transitions in the La–Ce–H system are considered to be arising from the main phase of $P6_3/mmc$-(La,Ce)H$_{9-10}$, rather than $C2/m$-LaH$_{10}$ [24]. Meanwhile, laser heating the sample at 152 GPa did not achieve as high $T_c$ as that of LaH$_{10}$. Previously, Geballe et al. have detected the evolution of the XRD patterns during the decompression of $Fm\overline{3}m$-LaH$_{10}$ from 169 GPa to 27 GPa (Fig. S8 in Ref. [9]). We noticed that the main peak abnormally shifted to the higher angles at 121–109 GPa, and some new peaks emerged, which was characteristic of the decomposition and appearance of a new phase, possibly $P6_3/mmc$-LaH$_x$. This is not in accord with the nearly linear $T_c$–$P$ trend of the current La–Ce–H system. Third, the synthesized compounds of the La–Ce–H system have much higher $T_c$ than the binary La–H and Ce–H phases in the same pressure range. For instance, the La–H phase was synthesized at a similar pressure with much lower $T_c$ of 112 K at 129 GPa, which decreased to 84 K at 118 GPa and then increased again. The figure of merit $S = \frac{T}{\sqrt{P^2+T'^2}}$ [48] has been proposed to indicate the quality of the various superconductors, where $T$ is the critical temperature, $P$ is the related pressure and $T'$ is the $T_c$ of MgB$_2$ at ambient pressure, equal to 39 K. As Fig. 4b shows, the alloying of La and Ce with initial ratio of 2.5–3.5:1 led to the discovery of ternary $hcp$-(La,Ce)H$_{9-10}$ with $T_c$ reaching 176 K at about 100 GPa and quite a high figure of merit $S = 1.62$ among the experimentally reported superconducting hydrides. Summing up, we not only synthesized two superconducting hydrides with $S > 1$ ($S_{MgB2} = 1$), but also enhanced the confidence to reach the next milestone like cuprates superconductors with much higher figure of merit ($S = 3.5$ for HgBaCaCuO).

**Conclusions**

In this work, we successfully synthesized the high-temperature superconducting ternary La–Ce–H (La: Ce = 2.5~3.5) and binary La–H compounds at pressures lower than 130 GPa. The synchrotron XRD data revealed that the new ternary La–Ce–H phase has the hexagonal structure with the hydrogen content reaching 9-10 per unit cell, i.e. *hcp*-(La,Ce)$H_{9-10}$. This compound showed a relatively linear $T_c$–$P$ tendency with the rate of about –0.6 K/GPa upon decompression from 130 to 105 GPa. Surprisingly, $T_c$ of *hcp*-(La,Ce)$H_{9-10}$ was preserved to 176 K at about 100 GPa, which meant a quite high figure of merit ($S$=1.62) among the synthesized hydrides. Besides, an extrapolation of the upper critical magnetic field $H_{c2}(0)$ gives 216 T at 100 GPa, the highest value ever observed for polyhydrides. $T_c$ of the binary La–H phase was 112 K at 129 GPa and decreased first to 84 K at 118 GPa, then increased again to above 100 K at 78 GPa. In the pressure range of 95–130 GPa, the ternary hexagonal La–Ce–H system shows higher $T_c$ than the binary La–H system, with the maximum difference reaching 100 K. The compounds of both these systems were synthesized at the same pressure and temperature conditions. This study indicates the prospects for further research of alloy-based high-$T_c$ hydrides at moderate pressures.

**Acknowledgments**

The authors thank the staff of the Shanghai Synchrotron Radiation Facility for their help during the synchrotron XRD measurements. This work was supported by the National Key R&D Program of China (Grant No. 2018YFA0305900), the National Natural Science Foundation of China (Grants No. 52072188, No. 11974133, No. 51632002, and No. 51720105007), and the Program for Changjiang Scholars and Innovative Research Team in University (Grant No. IRT_15R23). D.V.S. thanks the Russian Foundation for Basic Research (project 20-32-90099) and the Russian Science Foundation, grant 22-22-00570. A.R.O. thanks the Russian Science Foundation (grant 19-72-30043).

# Supplemental Information

## for

## Enhancement of the superconducting critical temperature realized in the La–Ce–H system at moderate pressures

Wuhao Chen,[1] Xiaoli Huang,[1,*] Dmitrii V. Semenok,[2] Su Chen,[1] Kexin Zhang,[1] Artem R. Oganov[2] and Tian Cui[3,1,*]

[1] State Key Laboratory of Superhard Materials, College of Physics, Jilin University, Changchun 130012, China
[2] Skolkovo Institute of Science and Technology, Skolkovo Innovation Center, Bolshoy Boulevard 30, bldg. 1, Moscow 121205, Russia
[3] School of Physical Science and Technology, Ningbo University, Ningbo 315211, China

*Corresponding authors' e-mails: huangxiaoli@jlu.edu.cn, cuitian@nbu.edu.cn

**Experimental details**

The La–Ce alloys were prepared using the multitarget magnetron sputtering (Fig. S1). We sputtered the La (99.9%) and Ce (99.9%) metals simultaneously to the glass slide using DC and RF power supplies, respectively. The Ar pressure was 1.5 Pa, and both targets were pre-sputtered to remove the surface oxides. Depending on the distance to the target, the La: Ce ratio has a certain distribution. The La–Ce alloy was further characterized using the scanning electron microscope (SEM) Regulus 8100, equipped for the energy dispersive X-ray spectroscopy (EDX). The La–Ce alloys for high pressure experiment were located on the glass side very close to the SEM-characterized area and kept strictly in the glove box ($O_2$ < 0.01 ppm, $H_2O$ < 0.01 ppm). Normal type-Ia diamonds with 60–150 μm culets single-beveled to 250–300 μm were used; the pressure was measured according to the Raman vibration edge of diamond using Akahama's calibration[1,2]. The resistance was measured using the four-probe method with the delta model of the Keithley current source (Model 6221, 1 mA) and voltmeter (Model 2182A). The electrodes were integrated to the diamond by a lithographic Mo (300-500 nm thick) [3] or manually cutting Pt (2–3 um thick) foil. To protect the electrodes during decompression, we tested the nano-polycrystalline diamonds (NPD) without bevels in some runs. Because of the highly fluorescence background of the NPDs, their Raman edge cannot be distinguished. Therefore, the NPDs (200–300 um culet) were combined with the normal diamonds, and the electrodes were set on the NPD side. The indentation of a tungsten gasket was insulated using c-BN/epoxy first, and the bevel part was filled with oxides ($Al_2O_3$, MgO)/epoxy. The La–Ce samples were loaded into the chamber filled with ammonia borane (AB) which acted as a hydrogen source. The sample was heated at the target pressure by a 1070 nm infrared laser with 3–5 um focus point and an exposure time of 1–3 s. After that, we put the DAC into a helium cryostat (1.5–300 K) equipped with a 0–9 T superconducting magnet for low-temperature electrical measurements. The crystal structure was determined using the synchrotron X-ray diffraction (XRD) on the BL15U1 synchrotron beamline with a wavelength of 0.6199 Å at the Shanghai Synchrotron Research Facility (SSRF). The experimental XRD images were integrated and analyzed for possible phases using the Dioptas software package [4]. To fit the diffraction patterns and obtain the cell parameter, we analyzed the data using Materials Studio and Jana2006 software [5] employing the Le Bail method[6].

Table S1. Initial samples and the pressure before and after the laser-heating in different experimental runs.

| Run | Initial sample | La:Ce | Pressure before and after laser heating | Measurements |
|---|---|---|---|---|
| #1 | (La,Ce)+AB | Unknown | 102–82 GPa | XRD |
| #2 | (La,Ce)+AB | Unknown | 113–112 GPa | XRD, R–T |
| #3 | (La,Ce)+AB | Unknown | 130–131 GPa | XRD, R–T |
| #4 | (La,Ce)+AB | ≈6 | 107–103 GPa | EDX, R–T |
| #5 | (La,Ce)+AB | 3.2–3.5 | 129–132 GPa | EDX, R–T |
| #6 | (La,Ce)+AB | 2.7–2.9 | 152–150 GPa | EDX, R–T |
| #7 | (La,Ce)+AB | Unknown | without laser-heating | R–T |
| #8 | (La,Ce)+AB | ≈3 | 129–131 GPa, 131–130 GPa | EDX, R–T |
| #9 | (La,Ce)+AB | 3.4–3.5 | 120–118 GPa, 122–120 GPa, 127–125 GPa | EDX, R–T |
| #L1 | La+AB | Pure La | 125–123 GPa, 130–127 GPa, 132–129 GPa | R–T |
| #L2 | La+AB | Pure La | 113–111 GPa | R–T |
| #L3 | La+AB | Pure La | 109 GPa heated | R–T |
| #S | La+AB | Pure La | 132–130 GPa | XRD |

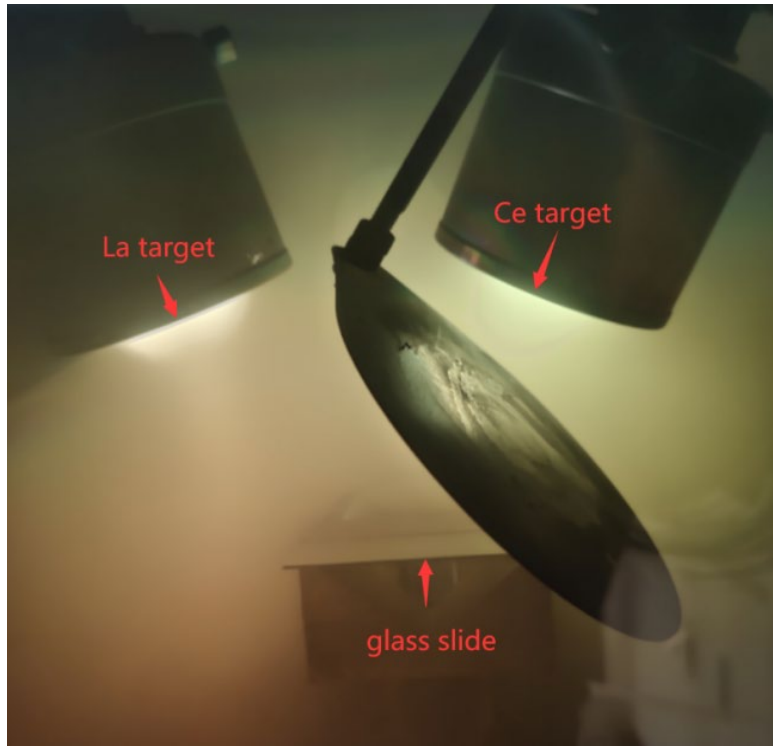

**FIG. S1.** Photograph of the chamber inside the magnetron sputtering equipment. The La and Ce ions were mixed in the Ar atmosphere and then deposited on a glass slide.

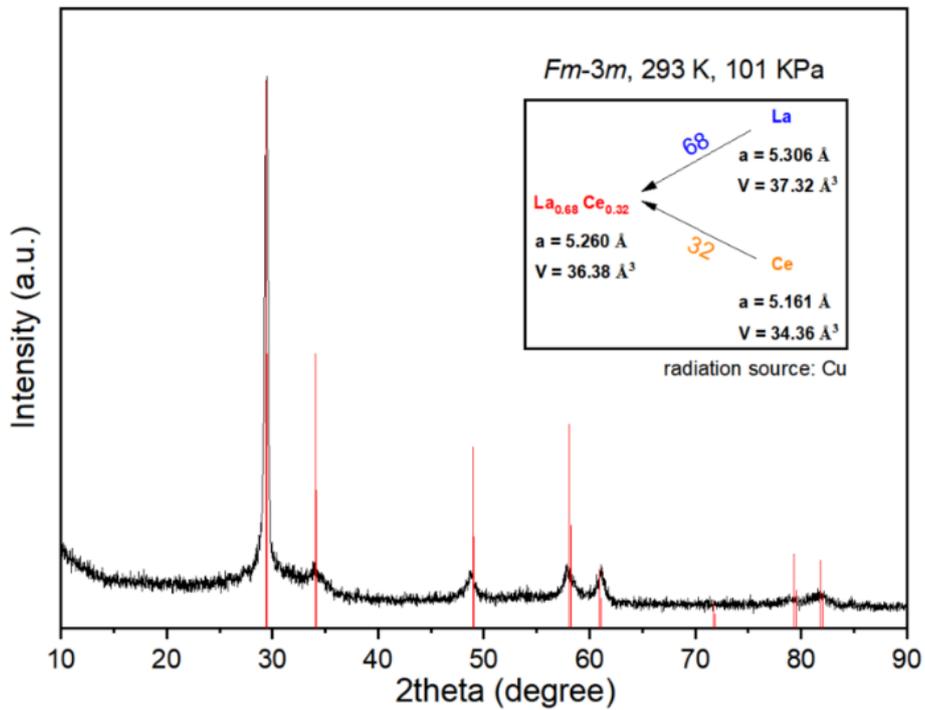

**FIG. S2.** XRD (Cu: $K_{\alpha 1}$=1.54056Å, $K_{\alpha 2}$=1.54439Å) of a typically prepared La–Ce alloy at ambient conditions. Red lines show the indexing results of the calculated diffraction of $Fm\bar{3}m$ structure. The related cell parameters are shown in the inset. The La–Ce ratio was estimated from the calculations of cell volume. The La contents decreases to 64% according to the $a$-spacing because of the deviation from the linear relation [7].

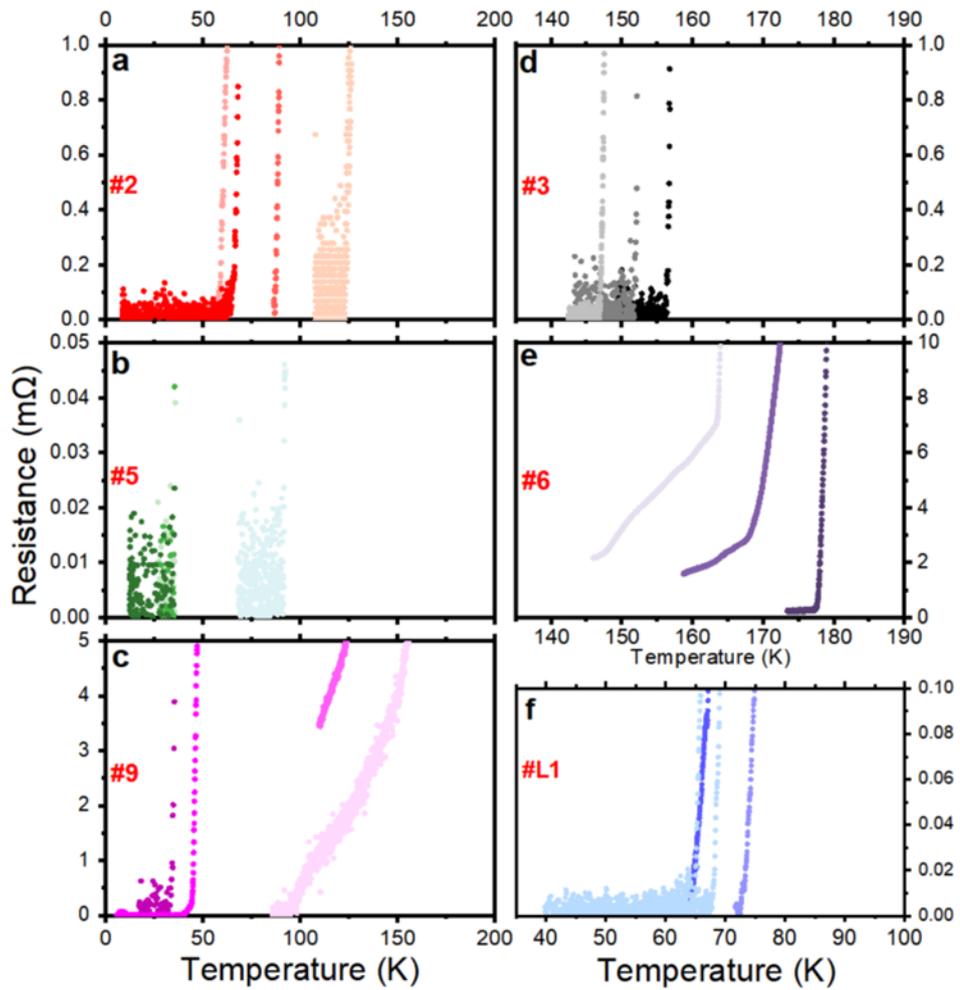

**FIG. S3.** Residual resistance of the samples (enlarged parts of Fig. 1 from the main text).

**Run #1**

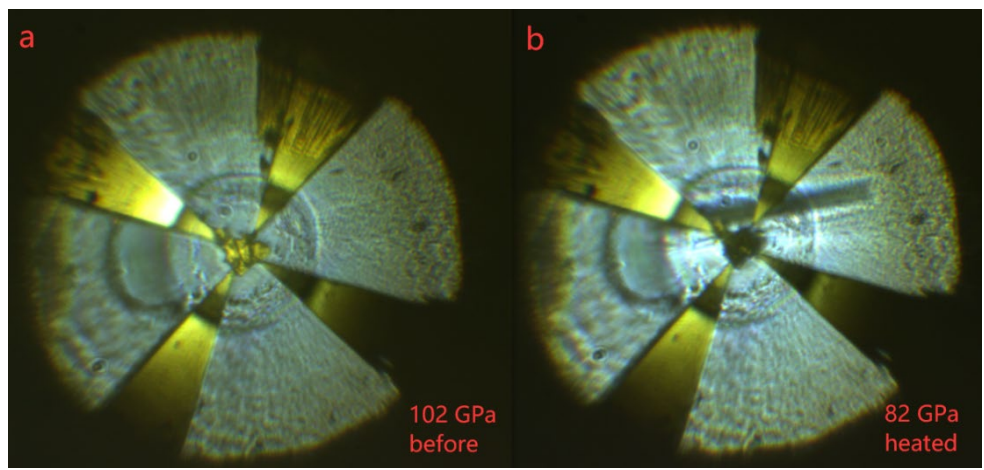

**FIG. S4.** Photographs of DAC #1 (a) before and (b) after laser-heating. A vertical crack appeared and grew gradually during the laser-heating.

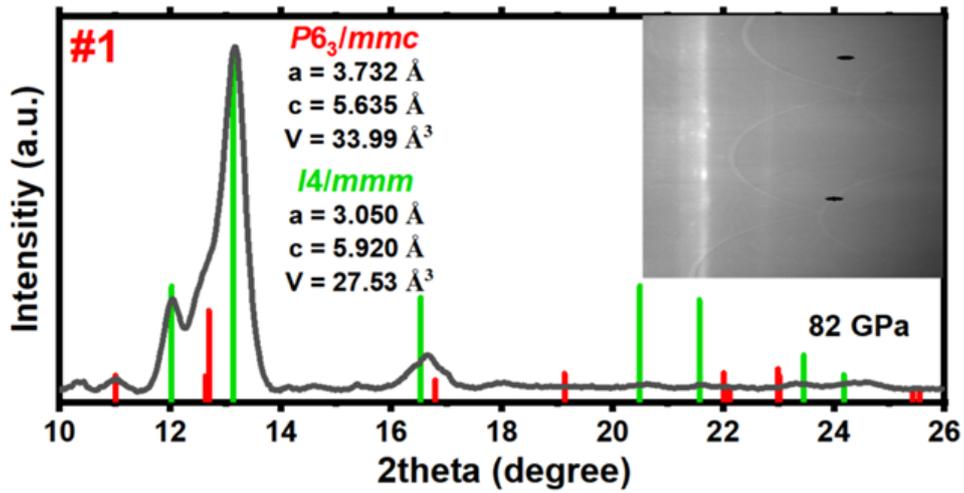

**FIG. S5.** Le Bail fitting of the XRD pattern of the sample in DAC #1 at 82 GPa. The inset shows the integrated XRD pattern.

Run #2

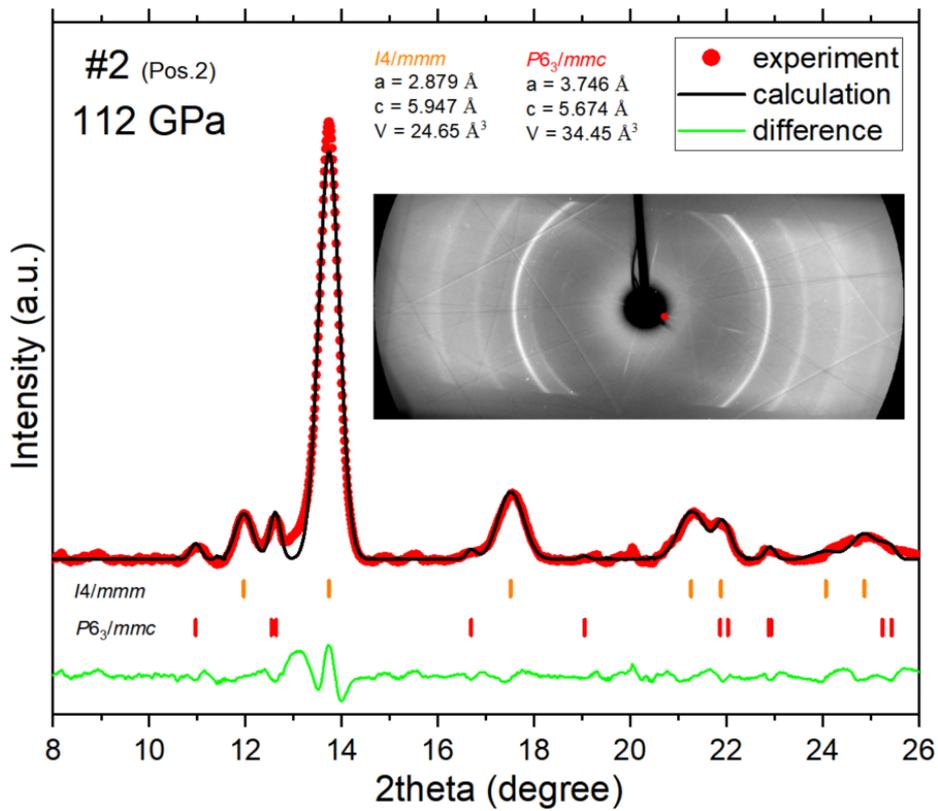

**FIG. S6.** Le Bail fitting of the XRD pattern for another detected position of DAC #2 at 112 GPa. The inset shows the XRD pattern.

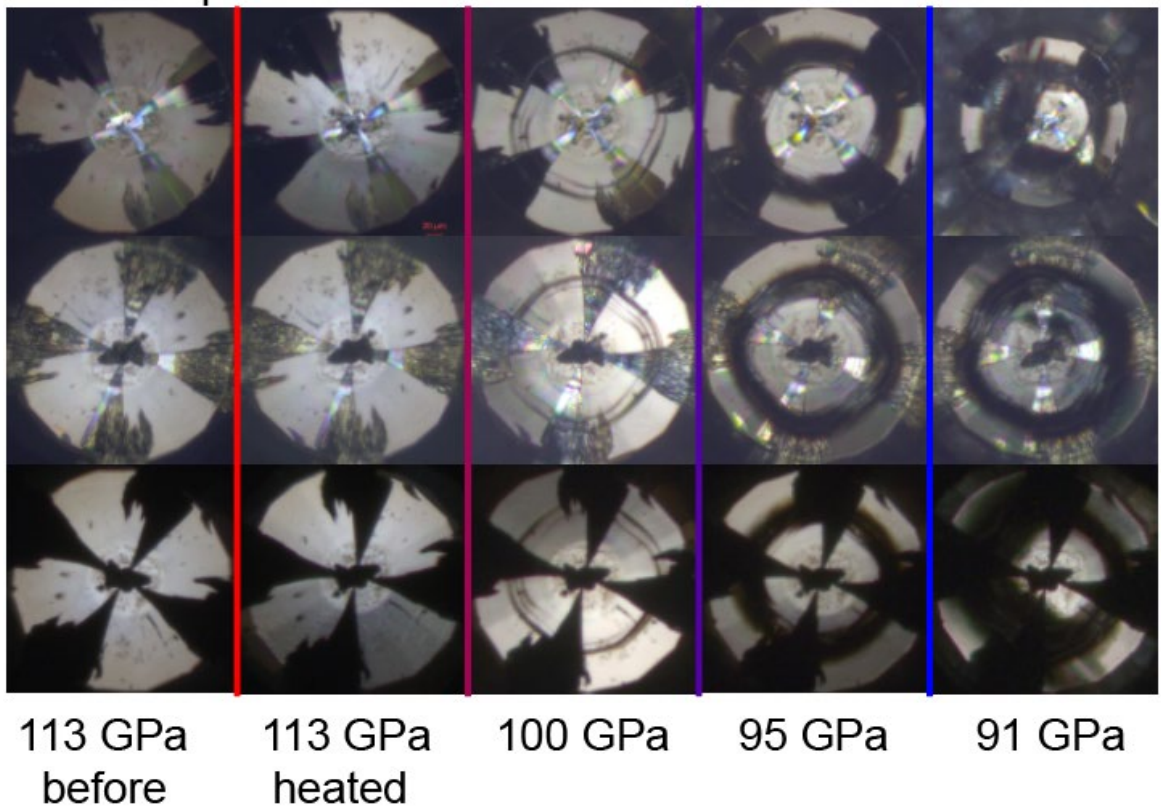

**FIG. S7** Changes in the diamond's bevel during the decompression of DAC #2 from different photographic views.

**Run #3**

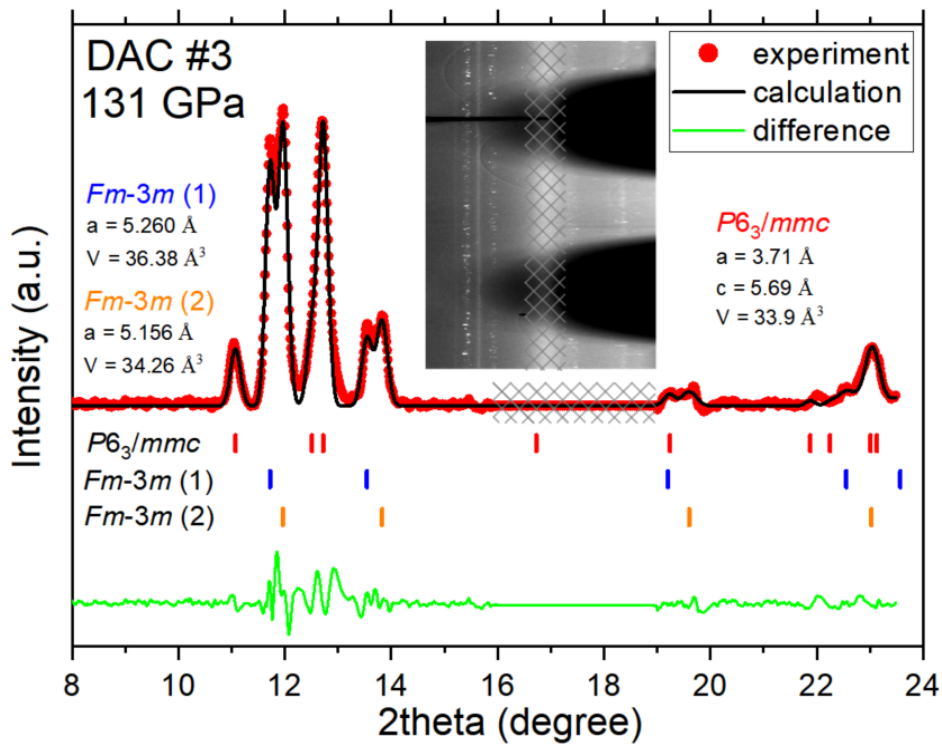

**FIG. S8.** Le Bail fitting of the XRD pattern of the sample in DAC #3 at 131 GPa. The inset shows the integrated diffraction pattern. The diffraction on the impurity that located on the seat surface of the diamond, as discussed below, is masked by gridlines.

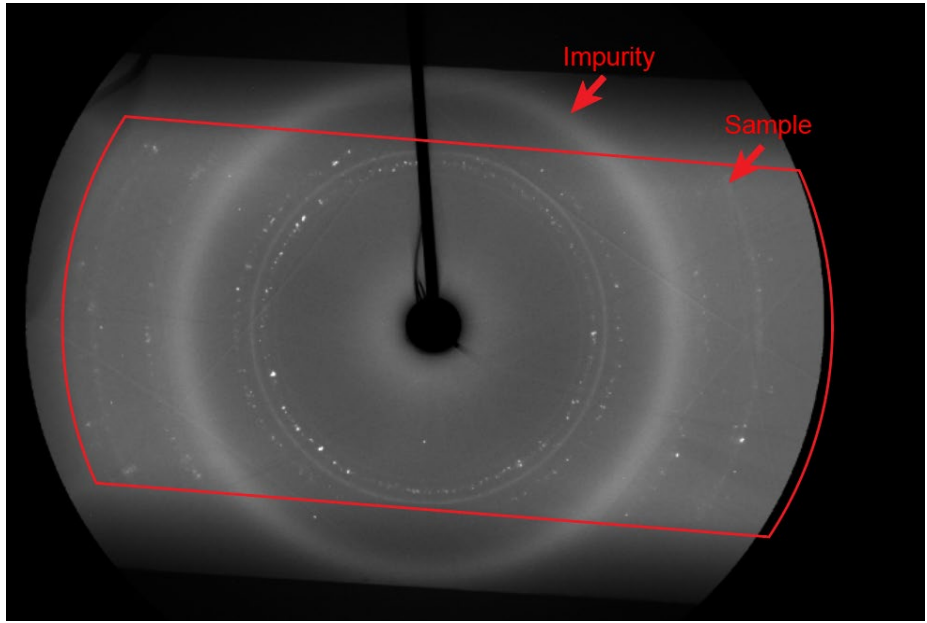

**FIG. S9.** XRD diffraction pattern of the sample in DAC #3 at 131 GPa. The red frame represents the projection of the long opening angle in the diamond seat. The sample shows a set of incomplete diffraction circles because of the occlusion of the seat, whereas the impurity shows a full broad entire circle. This means the impurity is not in the sample chamber and is closer to the diamond seat (possibly some dirt on the seat surface of diamond).

**Run #4**

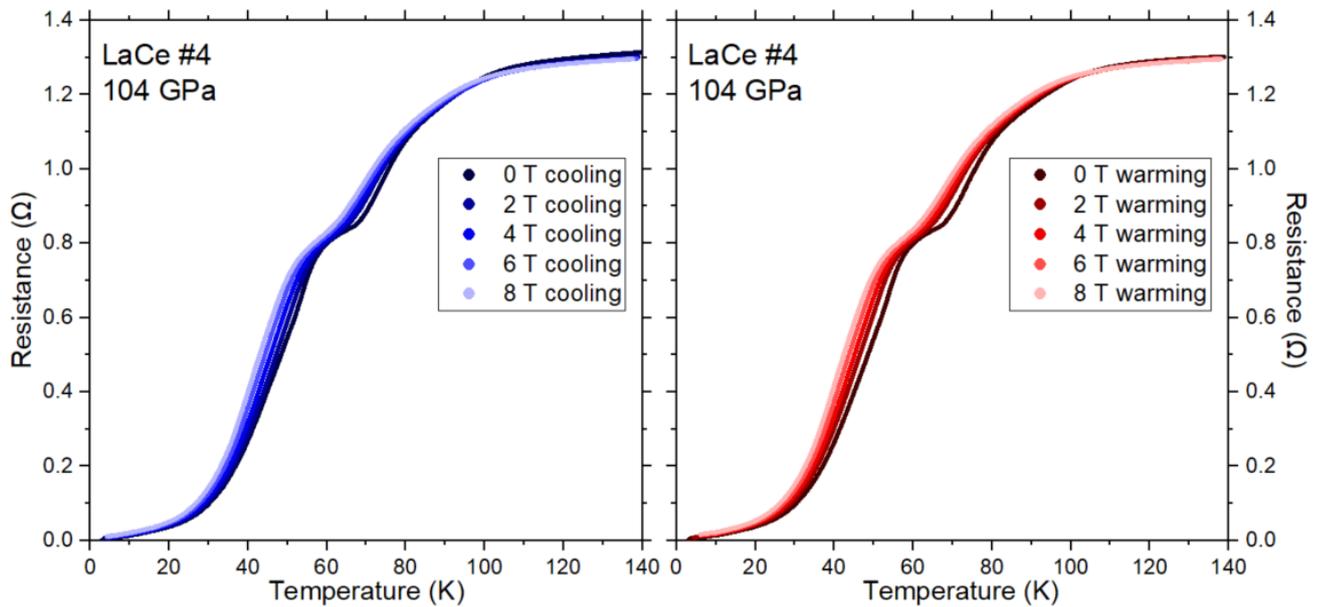

**FIG. S10.** Temperature dependence of the electrical resistance in DAC #4 near the superconducting transition in an external magnetic field for cooling(left) and warming(right) cycles.

**Run #5**

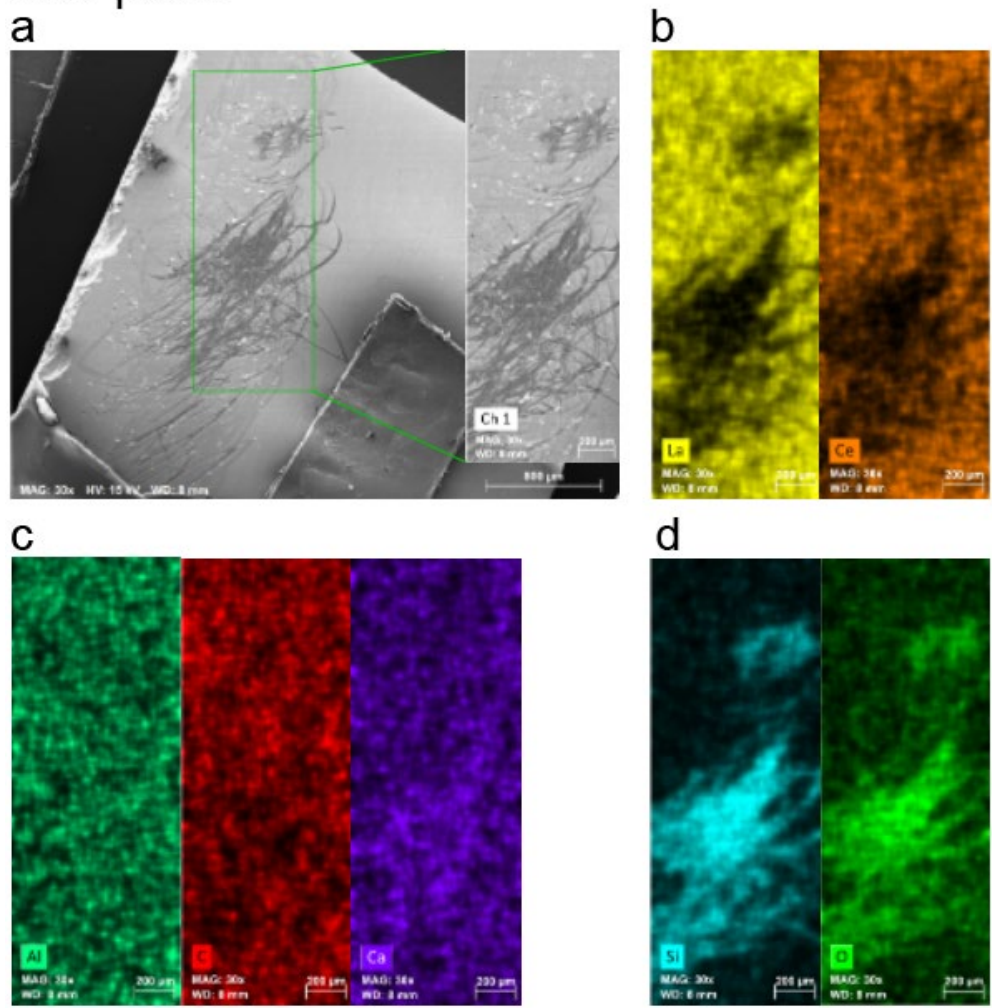

**FIG. S11.** Scanning electron microscopy (SEM) and energy dispersive X-ray spectroscopy (EDX) analysis of the La–Ce alloy in DAC #5, pos. 1 (a) The SEM photo of the sample on the glass slide. (b–d) Elements distribution in the rectangular area shown by green lines in panel (a).

In DAC #5, the alloy composition was characterized after the sample loading. We used a tungsten needle to scratch the La–Ce layer on the glass slide to get particles, which left traces (Fig. S11a). La and Ce were both uniformly distributed except for scratched area (Fig. S11b). Si and O mainly come from the glass slide, thus the scratched area has high intensity (Fig. S11d). On the La–Ce surface, there was a tiny amount of oxygen because of the oxides formed during the transfer. Moreover, Al, C and Ca exist all over the selected region and can be viewed as a background signal. Fig. S13 shows that the scratched particles have a uniform distribution of La and Ce. The sputtered La–Ce can form clusters or smooth solid solution.

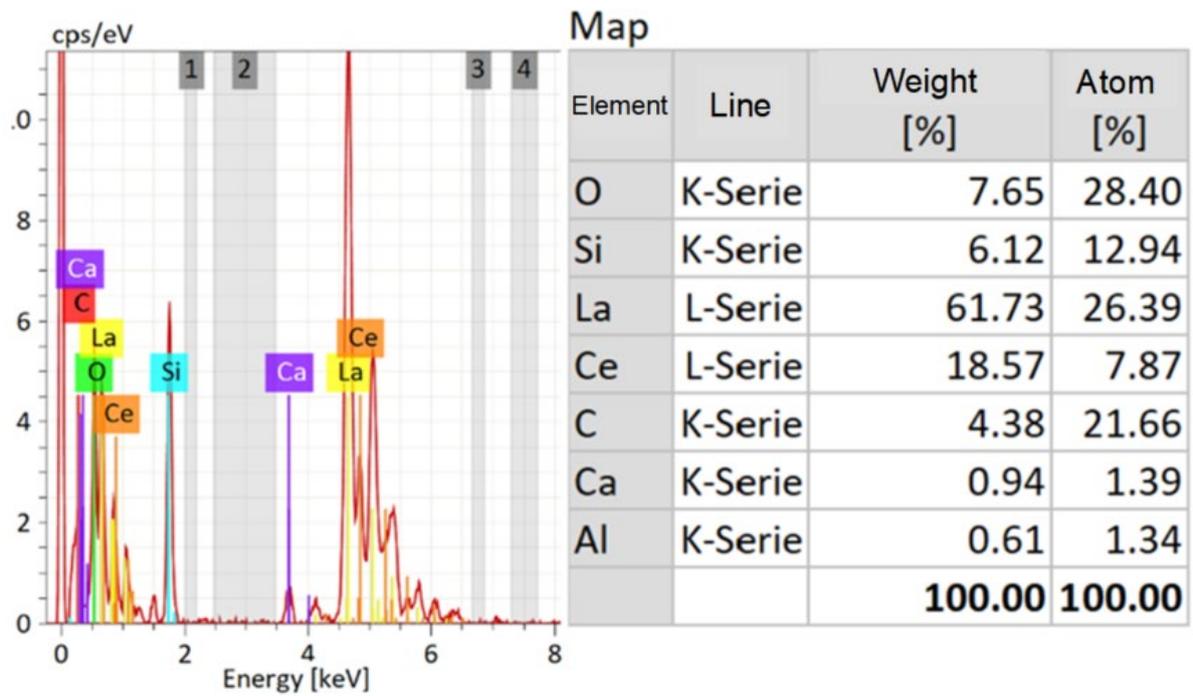

**FIG. S12.** Compositions analyzed using the EDX in DAC #5.

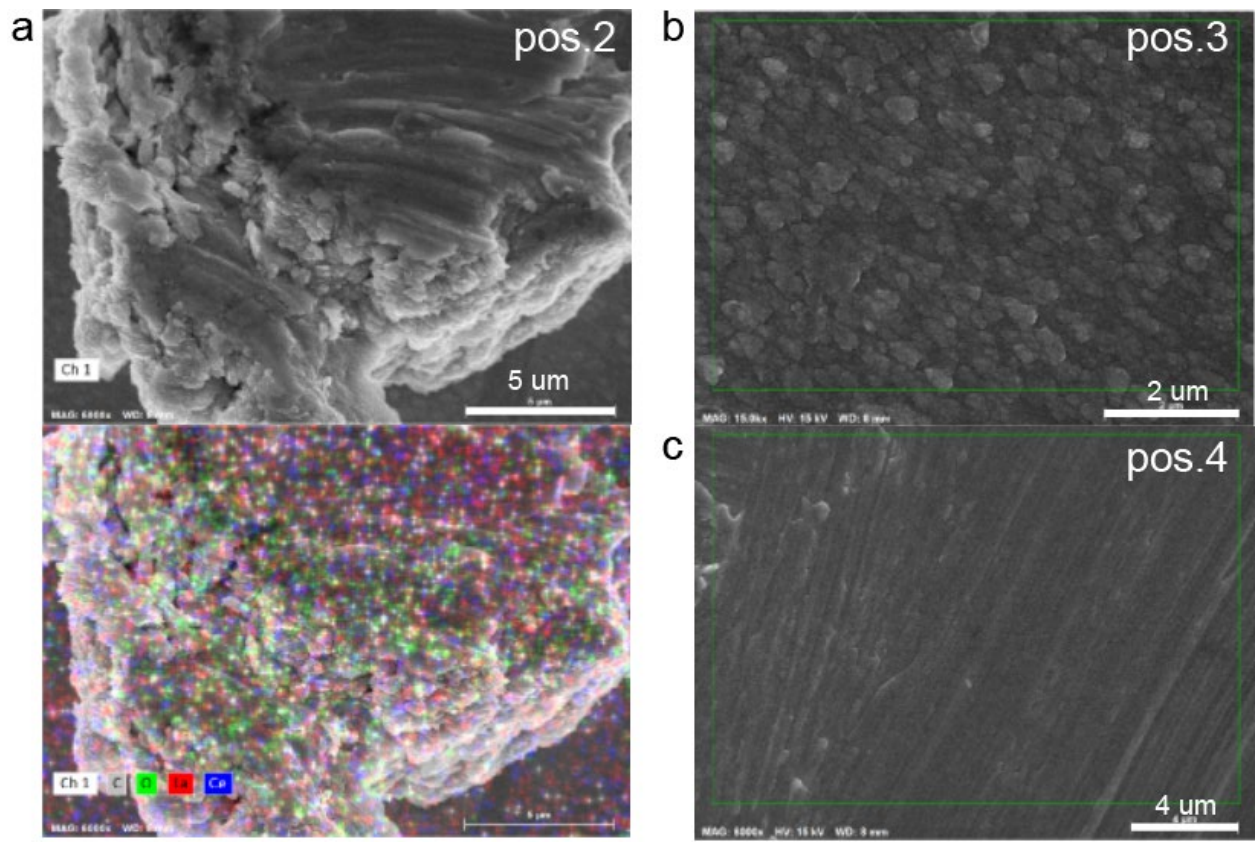

**FIG. S13.** The SEM photos of other positions in DAC #5. (a) The selected particle (top) and the element distribution on it (bottom). (b, c) Positions 3 and 4.

**Run #7**

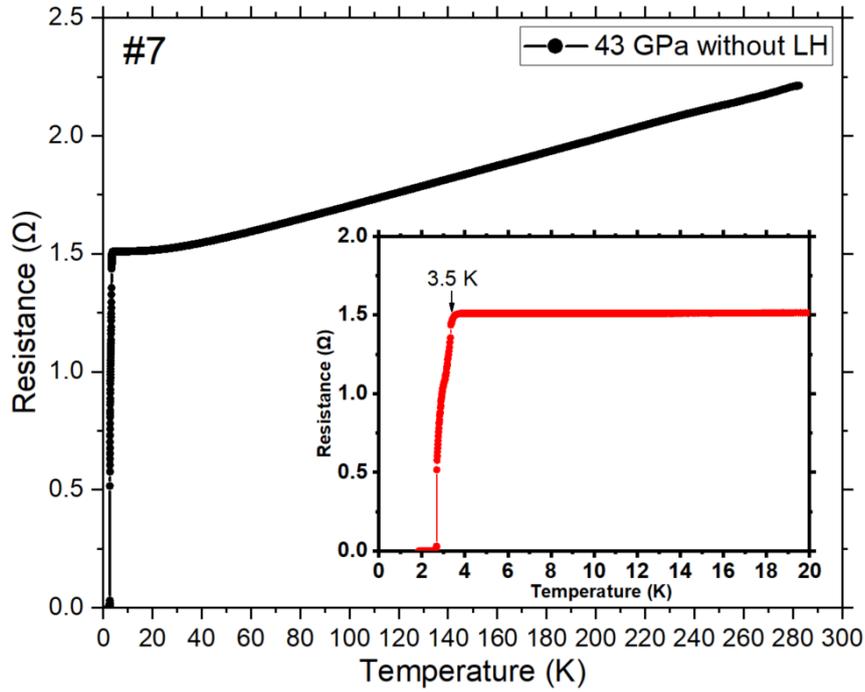

**FIG. S14.** $R$–$T$ dependence of La–Ce alloy in run #7 at 43 GPa before the laser-heating (LH). Inset shows the enlarged part of the plot.

**Run #8**

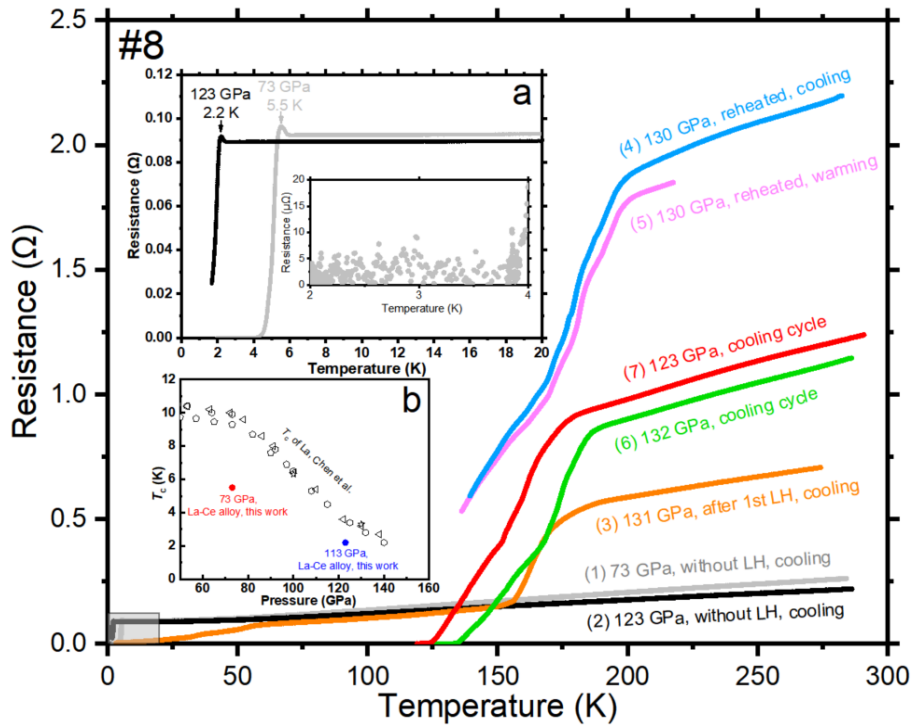

**FIG. S15.** $R$–$T$ dependence of the sample in DAC #8 at different conditions. Numbers from (1) to (7) indicate the sequence of the experiments. Inset (a) shows the enlarged part of the plot, with the superconducting transition of La–Ce alloy before the laser-heating. In inset (b), a comparison of $T_c$ of the La–Ce alloy and pure La is shown [8].

**Run #9**

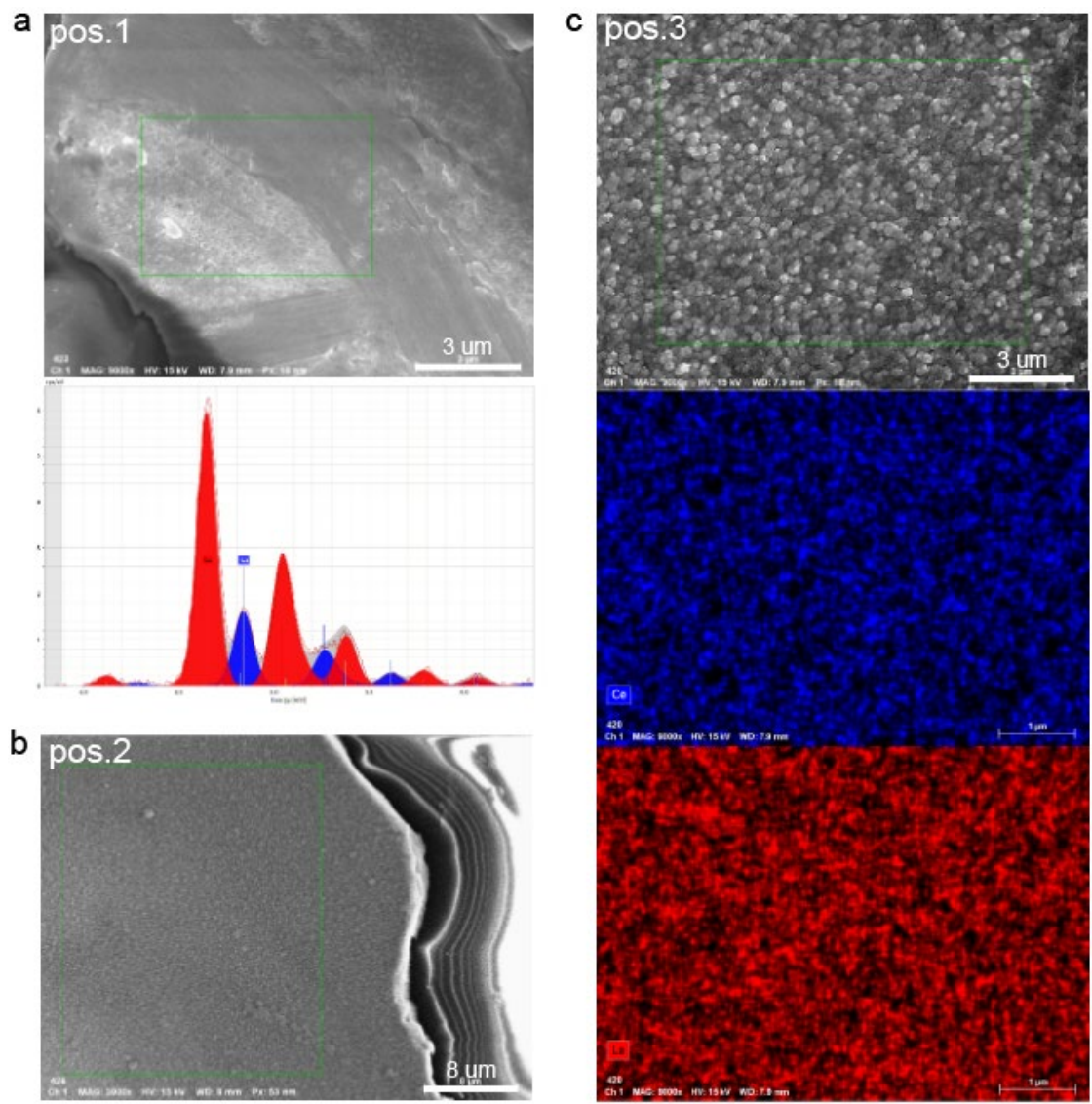

**FIG. S16.** The EDX analysis of the La–Ce alloy in DAC #9. (a) Top: the selected particle, bottom: L-Serie energy spectrum of the La–Ce alloy. (b) SEM photo at position 2. (c) Top: SEM photo at position 3, middle: Ce distribution, bottom: La distribution

**Run #L1**

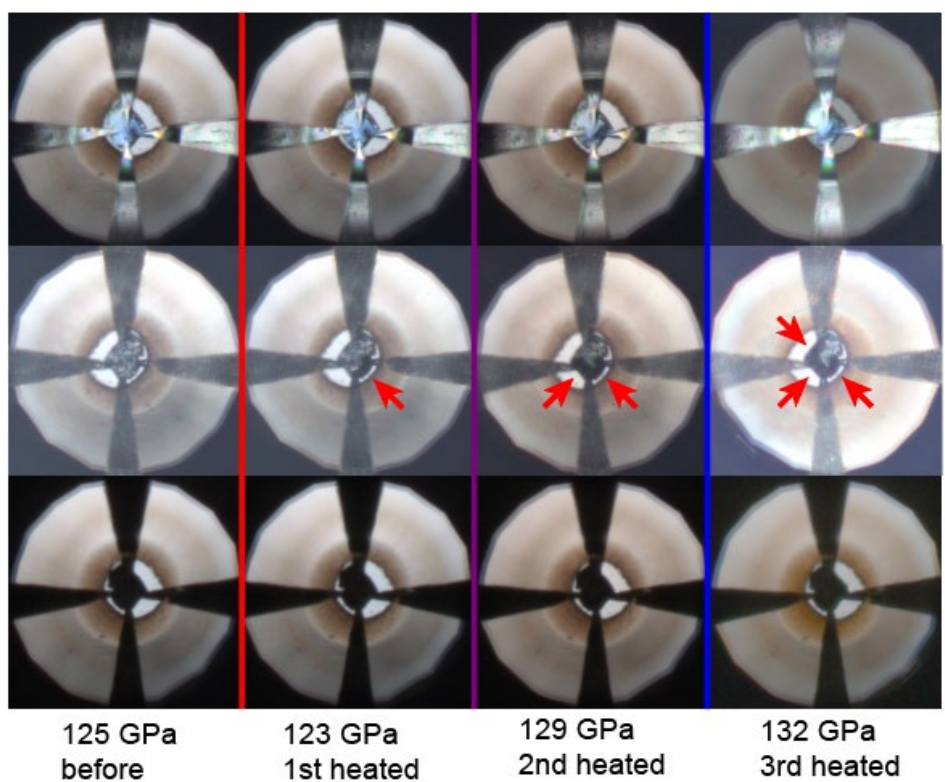

**FIG. S17.** Photographs of the sample in DAC #L1 before and after laser-heating for three times. The tips of four Pt electrodes were directly compressed to the edge of the sample. Arrows show the changes of the sample after the heating.

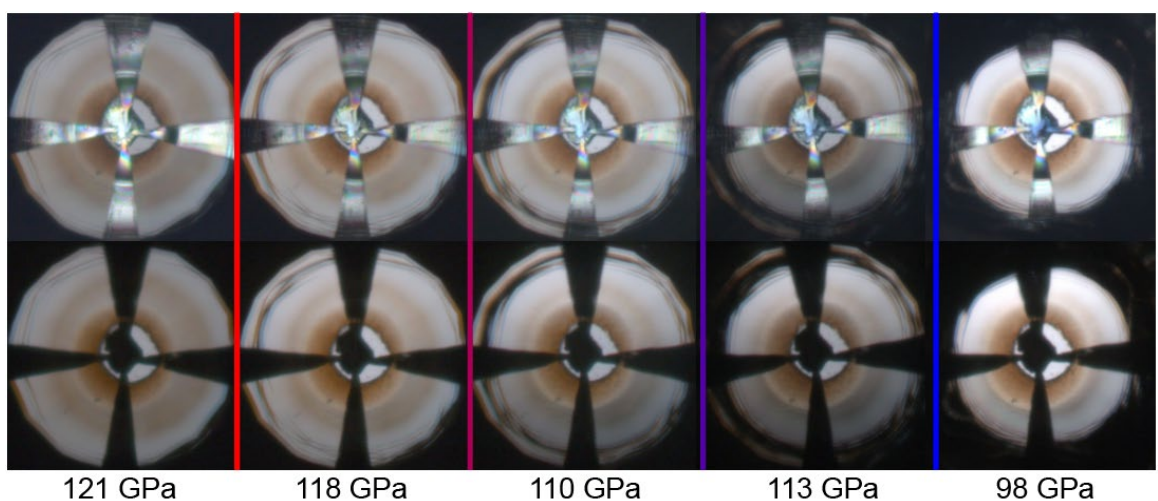

**FIG. S18.** Changes of the diamond's bevel during decompression of DAC #L1 at different pressures.

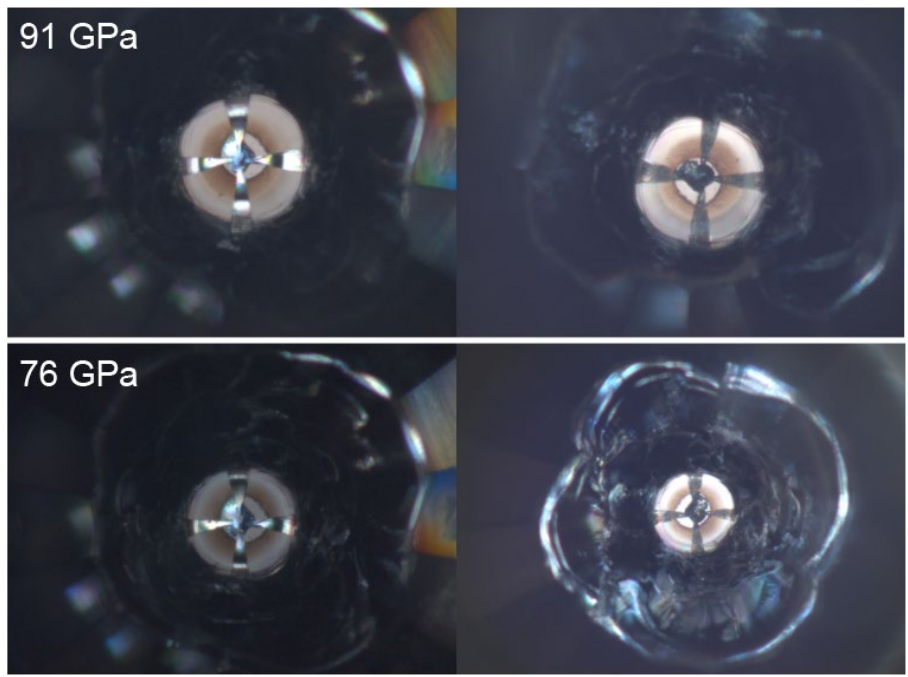

**FIG. S19.** Front and back views of the diamonds in DAC #L1 at 91 GPa and 76 GPa.

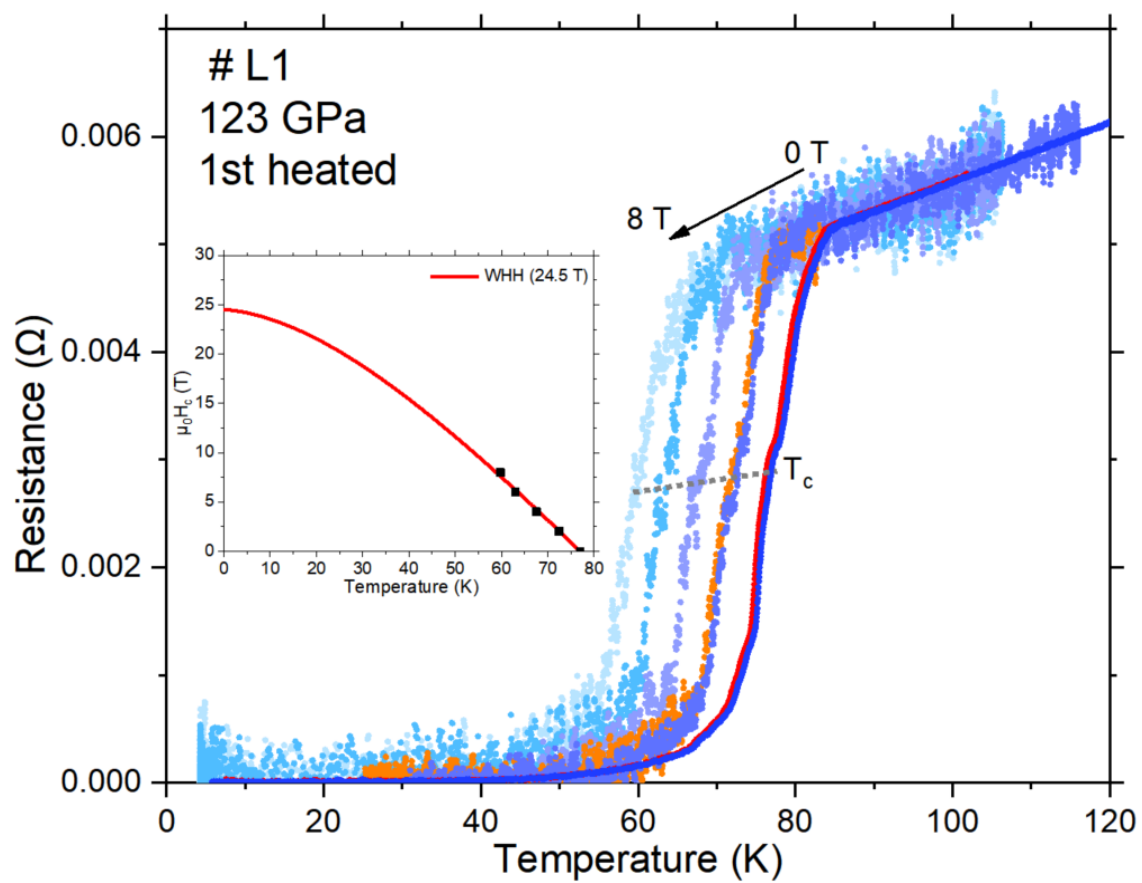

**FIG. S20.** Superconducting transition in DAC #L1 characterized by the temperature dependence of the resistance in an external magnetic field at 123 GPa after the first laser-heating. Inset shows the WHH fitting of the experimental data.

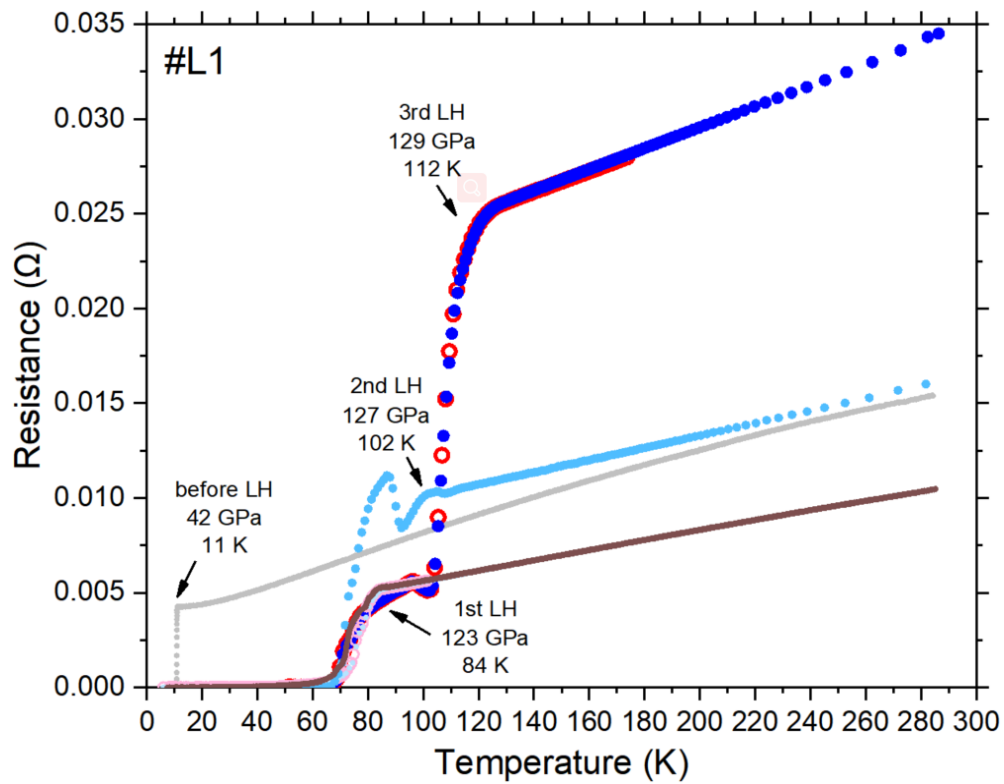

**FIG. S21.** Temperature dependence of the electrical resistance after laser-heating (LH) for three times. Solid circles represent the cooling cycle, and open circles represent the warming cycle. The horizontal density of the data points indicates the rate of the temperature change.

**Run #L2**

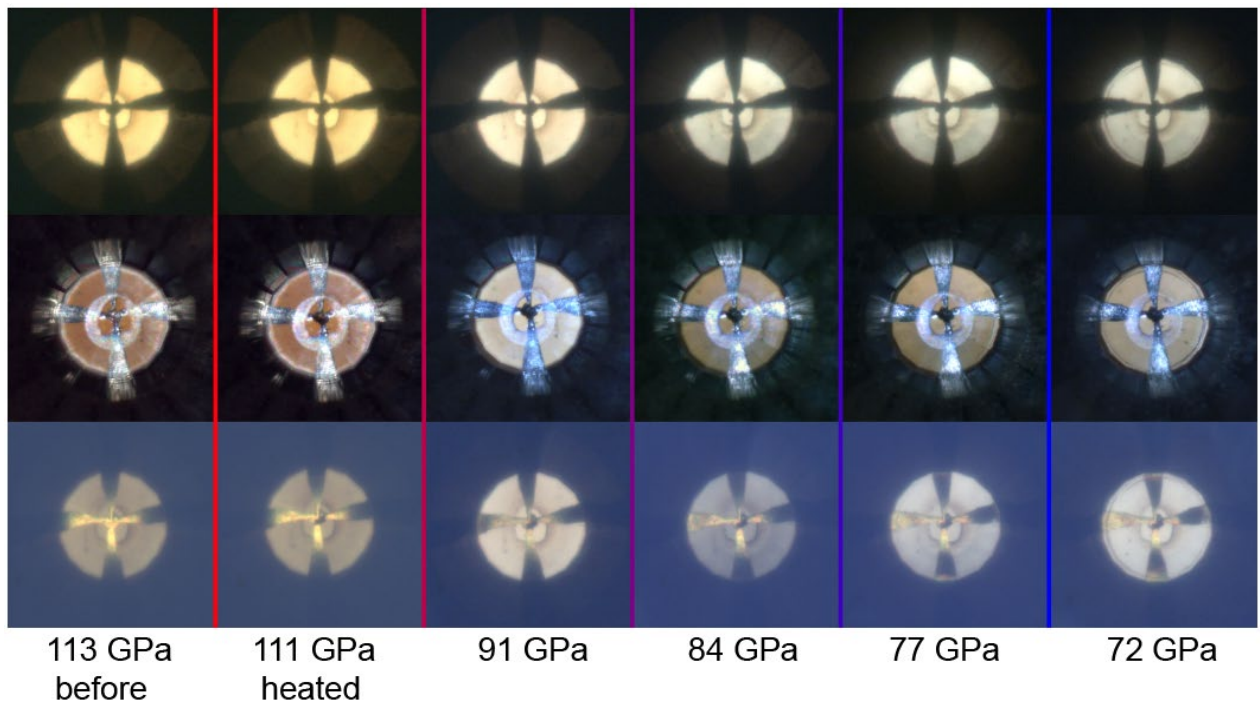

**FIG. S22.** Different photographic views of the diamond's bevel during the decompression of DAC #L2.

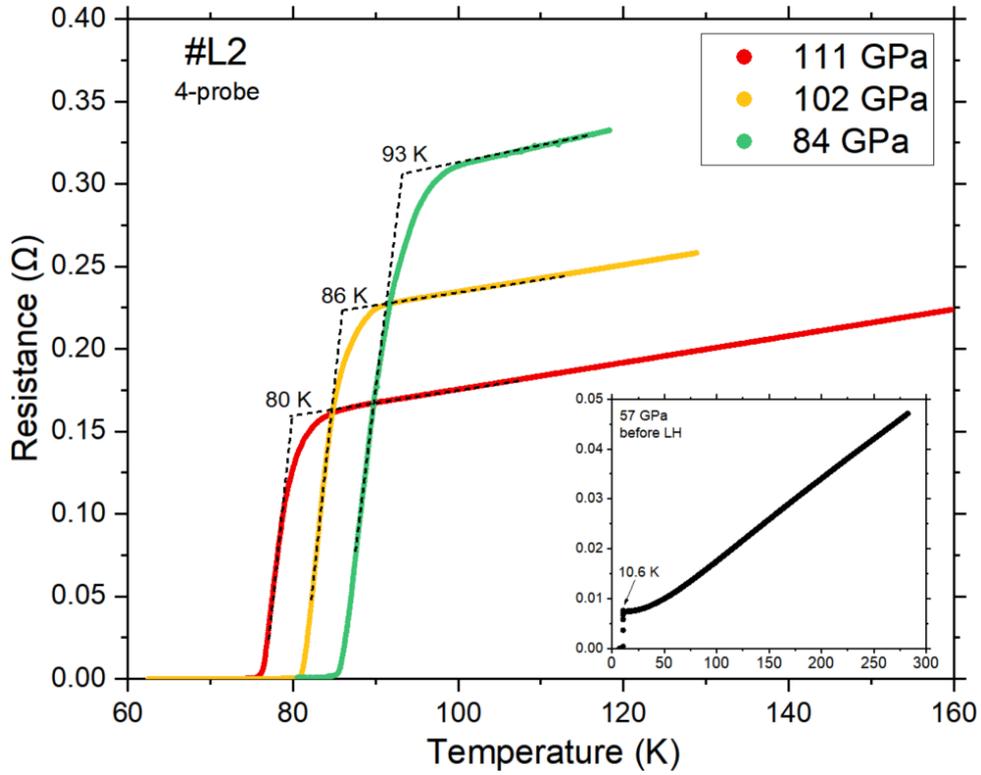

**FIG. S23.** Temperature dependence of the electrical resistance, measured using the four-probe method, showing the superconducting transitions of $LaH_x$ in DAC #L2. Inset shows the superconducting transition of pure La at 57 GPa before the laser-heating.

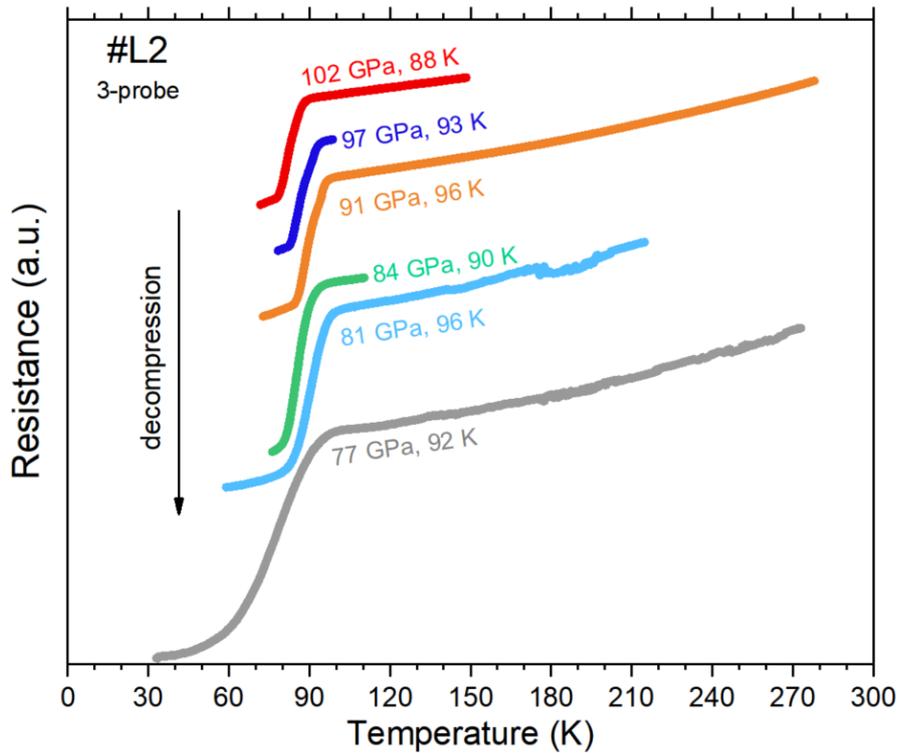

**FIG. S24.** Temperature dependence of the electrical resistance, measured using the three-probe method, showing the superconducting transitions of $LaH_x$ in DAC #L2 during decompression.

**Run #L3**

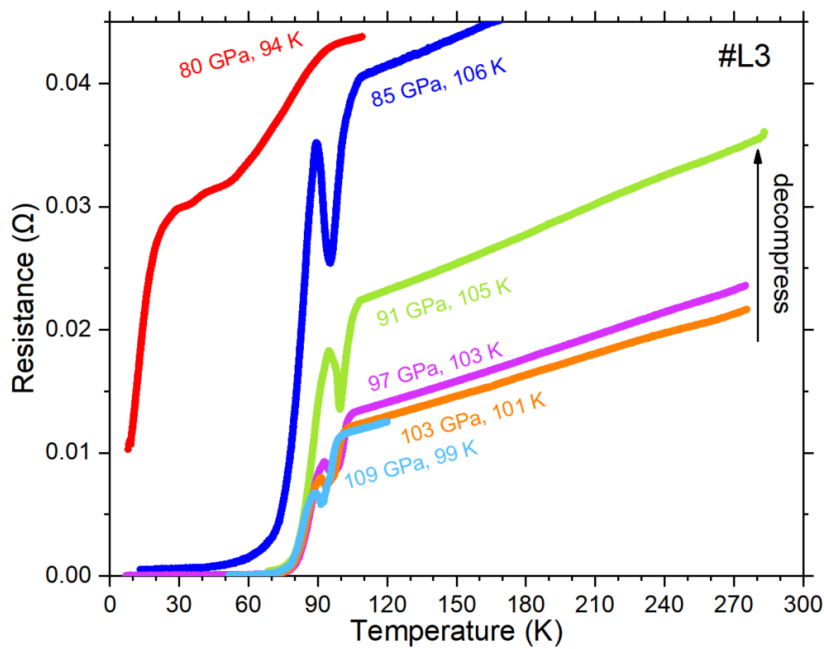

**FIG. S25.** Temperature dependence of the electrical resistance showing the superconducting transitions of LaH$_x$ in DAC #L3.

**Run #S**

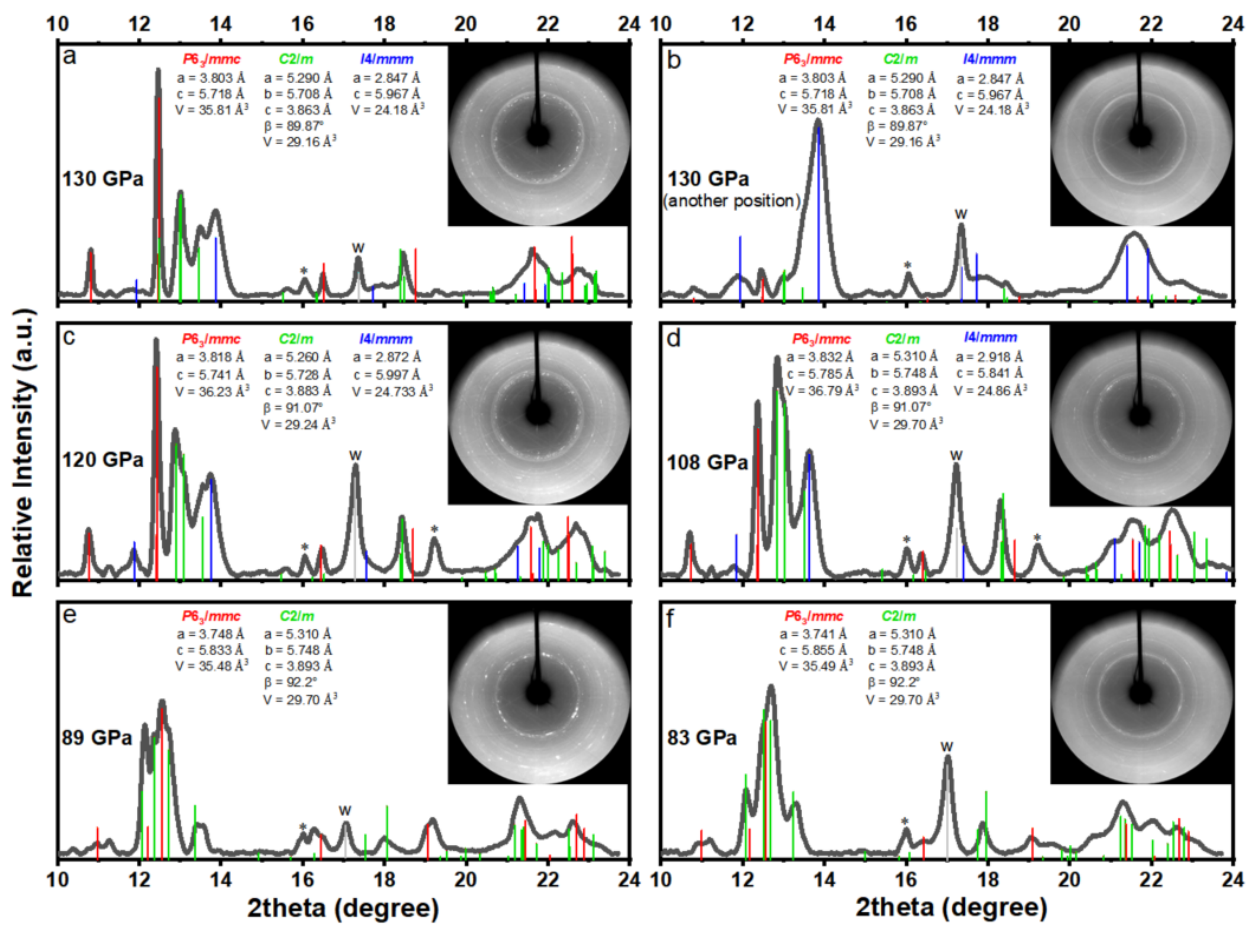

**Fig. S26.** Indexing of the synchrotron XRD data ($\lambda$ = 0.6199 Å) for the sample in DAC #S during decompression. Insets show the XRD patterns.